\documentclass[aps,prl,twocolumn,preprintnumbers,superscriptaddress,amsmath]{revtex4}

\usepackage{amssymb}
\usepackage{txfonts}
\usepackage{amssymb}
\usepackage{graphicx}
\usepackage{epstopdf}
\usepackage{subfigure}
\usepackage{dcolumn}
\usepackage{bm}
\usepackage{color}
\usepackage{upgreek}

\begin{document}

\title{Generation of a frequency-degenerate four-photon entangled state using a silicon nanowire}

\author{Lan-Tian Feng}
\affiliation
{Key Laboratory of Quantum Information, University of Science and Technology of China, CAS, Hefei, 230026, People's Republic of China.}
\affiliation{Synergetic Innovation Center of Quantum Information $\&$ Quantum
Physics, University of Science and Technology of China, Hefei, Anhui
230026, China.}
\author{Ming Zhang}
\affiliation{State Key Laboratory for Modern Optical Instrumentation, Centre for Optical and Electromagnetic Research,
Zhejiang Provincial Key Laboratory for Sensing Technologies, Zhejiang University, Zijingang Campus, Hangzhou
310058, China.}
\author{Zhi-Yuan Zhou}
\author{Yang Chen}
\affiliation
{Key Laboratory of Quantum Information, University of Science and Technology of China, CAS, Hefei, 230026, People's Republic of China.}
\affiliation{Synergetic Innovation Center of Quantum Information $\&$ Quantum
Physics, University of Science and Technology of China, Hefei, Anhui
230026, China.}
\author{Ming Li}
\author{Dao-Xin~Dai}
\affiliation{State Key Laboratory for Modern Optical Instrumentation, Centre for Optical and Electromagnetic Research,
Zhejiang Provincial Key Laboratory for Sensing Technologies, Zhejiang University, Zijingang Campus, Hangzhou
310058, China.}
\author{Hong-Liang Ren}
\affiliation{College of Information Engineering, Zhejiang University of Technology, Hangzhou 310023, P.R. China}
\author{Guo-Ping Guo}
\author{Guang-Can Guo}
\affiliation
{Key Laboratory of Quantum Information, University of Science and Technology of China, CAS, Hefei, 230026, People's Republic of China.}
\affiliation{Synergetic Innovation Center of Quantum Information $\&$ Quantum
Physics, University of Science and Technology of China, Hefei, Anhui
230026, China.}
\author{Mark Tame}
\affiliation
{Department of Physics, Stellenbosch University, Stellenbosch, South Africa.}
\author{Xi-Feng Ren\footnote[1]{renxf@ustc.edu.cn}}
\affiliation
{Key Laboratory of Quantum Information, University of Science and Technology of China, CAS, Hefei, 230026, People's Republic of China.}
\affiliation{Synergetic Innovation Center of Quantum Information $\&$ Quantum
Physics, University of Science and Technology of China, Hefei, Anhui
230026, China.}

\begin{abstract}
Integrated photonics is becoming an ideal platform for generating two-photon entangled states with high brightness, high stability and scalability. This high brightness and high quality of photon pair sources encourages researchers further to study and manipulate multi-photon entangled states. Here, we experimentally demonstrate frequency-degenerate four-photon entangled state generation based on a single silicon nanowire 1\ cm in length. The polarization encoded entangled states are generated with the help of a Sagnac loop using additional optical elements. The states are analyzed using quantum interference and state tomography techniques. As an example, we show that the generated quantum states can be used to achieve phase super-resolution. Our work provides a method for preparing indistinguishable multi-photon entangled states and realizing quantum algorithms in a compact on-chip setting.
\end{abstract}
\maketitle

\section*{1. INTRODUCTION}
Entangled quantum states are useful resources for many applications in quantum information science, including quantum teleportation~\cite{Bouwmeester1997}, one-way quantum computation~\cite{Raussendorf2001,Briegel2009}, quantum simulation~\cite{Lu2009,Georgescu2014} and quantum metrology~\cite{Giovannetti2006,Giovannetti2011}. Integrated photonics has long been recognized as a promising platform for realizing entangled quantum states due to a low pump power requirement, high stability, and scalability~\cite{Tanzilli2012,Caspani2017}, as well as showing advantages in portability for distributed quantum networks~\cite{Pers2013,McCutcheon2016,Bunandar2018}. A number of novel two-photon entangled states have been realized with integrated photonic circuits, such as high-dimensional entangled states~\cite{Kues2017,Wang2018} and spatial mode entangled states~\cite{Feng2016,Feng2018}. On the basis of high-brightness and high-quality two-photon sources, we can look further to the preparation of multi-photon quantum states. In this situation, manipulating non-degenerate multi-photon quantum states with a single-pump spontaneous four-wave mixing process (SFWM) is the most direct way~\cite{Reimer2016,Zhang2018}. However, the generated four-photon quantum states are still not entangled and the non-degenerate photon pairs have a very limited role in quantum information applications, especially in areas where there is a need for high-quality quantum interference with indistinguishable photons~\cite{Pan2012}.

A possible route to overcome the problem of degenerate multi-photon entangled states production is using two lasers with different frequencies to pump the nonlinear material in SFWM. This `dual-pump' technique was introduced in nonlinear integrated optics recently~\cite{Chen2006,Silverstone2014,He2015,Zhu2016,Li2018,Lin2007}, and has provided advantages over the single-pump process in quantum state manipulation~\cite{Lin2007,Fang2013,Christensen2016}. So far, there has been some experimental efforts made in producing correlated photon pairs with the dual-pump technique in integrated circuits~\cite{Silverstone2014,He2015}. However, experimental research on this approach is still scare at this time. Of particular importance is the generation of high-quality frequency-degenerate multi-photon entangled quantum states, which has not yet been proposed or studied in detail.

\begin{figure*}[t]
\centering
\includegraphics[width=16.0cm]{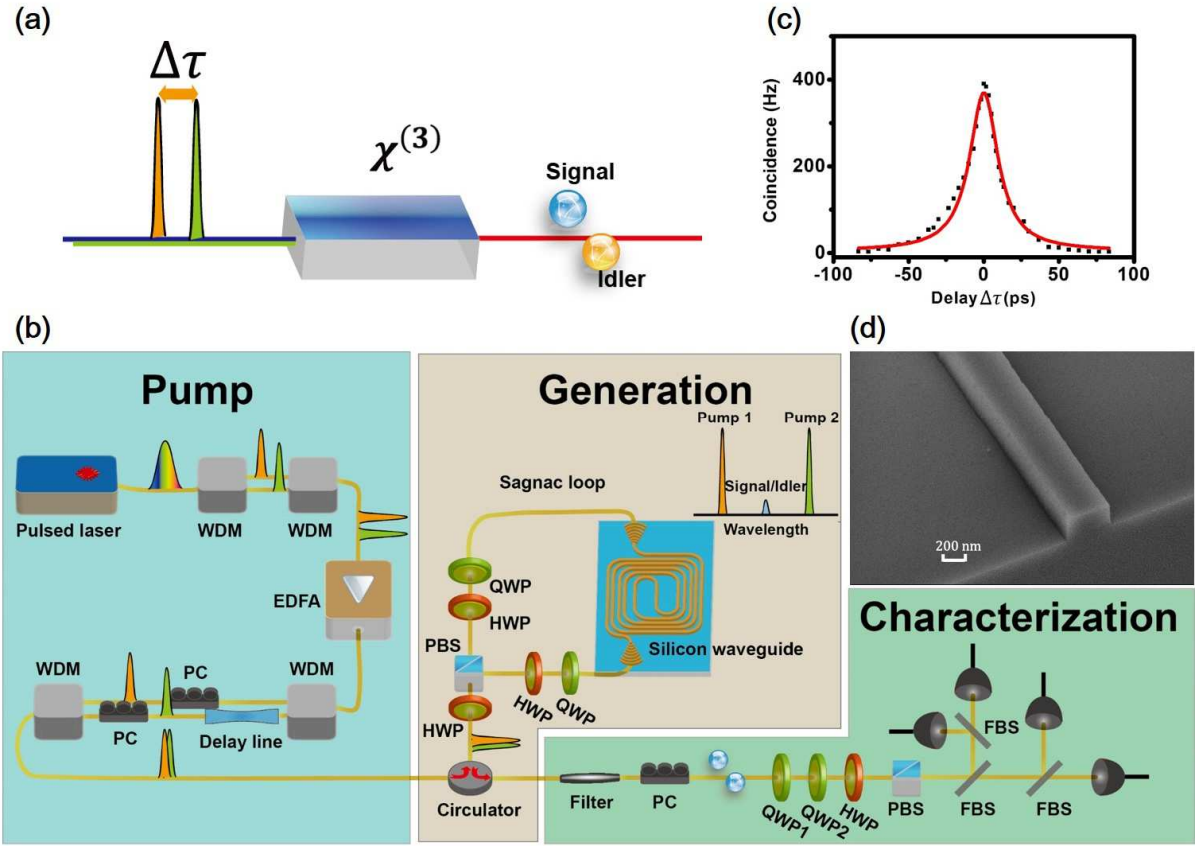}
 \caption {\textbf{Experimental generation of degenerate multi-photon quantum states.} \textbf{a}, Diagram showing the interactions in a third-order nonlinear medium with two pulsed pump lasers. \textbf{b}, Experimental setup for the simultaneous modulation of two pump pulses, degenerate quantum state generation and state characterization. In the region labeled `pump', a broadband pulsed laser was used as the pump source. After filtering and controlling the pulse with four wavelength-division-multiplexing (WDM) systems, the resulting two pulses with different frequencies were input into a Sagnac loop, in the region labeled `generation', for the generation of polarization encoded quantum states. In the region labeled `characterization', wave plates and fiber beam splitters (FBSs) constituted the measurement system for quantum state tomography, and thus to determine the quality of the generated quantum states. HWP: half-wave plate; QWP: quarter-wave plate; PC: polarization controller; PBS: polarization beam splitter; EDFA: erbium-doped fiber amplifier. \textbf{c}, Experimental results of the nonlinear interaction strength (leading to an increase in coincidences) by varying the time difference between the two pump pulses. \textbf{d}, SEM picture of the cross-section of the silicon waveguide.}
\end{figure*}

Here, by using dual-pump SFWM with a low pump power (several hundred $\upmu$W), we experimentally demonstrate frequency-degenerate four-photon entangled state generation based on two-photon state preparation with a silicon nanophotonic waveguide. The generated four-photon state is characterized by the tomography technique with a raw fidelity of $0.72\pm0.07$. We also demonstrate an application in quantum metrology for phase sensitivity measurements. Our approach could be further integrated and potentially used for preparing indistinguishable multi-photon entangled states and realizing quantum algorithms in a compact on-chip setting.

\section*{2. RESULTS}
\subsection{A. Experimental setup}

A diagram of the general scenario considered is shown in Fig. 1(a), where two pulsed lasers give rise to nonlinear interactions via a $\chi^{(3)}$ nonlinearity. When there is a time difference between the pulses, they will never interact because the pulse energy is highly localized in time. Therefore, only interactions in a single pulsed laser occur and non-degenerate photon pairs are generated from the single-pump SFWM. Precisely controlling the relative time delay of the two pulses is necessary for degenerate photon pair generation. Our experimental architecture to realize this precise control is shown in Fig. 1(b) in the region labeled `pump'. Here, a femtosecond erbium laser (T-light FC, pulse width $90$~fs) with a repetition rate of $100$~MHz is used as the pump source. The laser has a broad $3$~dB bandwidth covering the entire C band (about $80$~nm). We use four wavelength-division-multiplexing (WDM) systems to accomplish the filtering and controlling of two laser pulses. Each WDM has three channels, one for the total input/output signal, and the other two for separate output/input signals with different frequencies. Hence, two WDMs constitute one wavelength division and mixing system. The first two WDMs are used to achieve the selection of pulses with two central wavelengths of $1561.42$~nm and $1538.98$~nm from the pump source, each with a $50$~GHz $3$~dB bandwidth. Based on the calculation in ref. \cite{Zhang2018}, our silicon waveguide has a broad SFWM gain. In principle, we can choose any two pump pulses in this band and their center as our target photon-pair frequency. Here, we choose the pump wavelengths 1561.42\ nm and 1538.98\ nm as an example. Between the two WDMs, additional fiber filters (bandwidth $50$~GHz, extinction ratio $100$~dB) are used to purify the frequencies (not shown). It is worth pointing out that after the filters, the pulse width of the two pulses is no longer $90$~fs, and it is broadened to about $20$~ps (a brief calculation is given in Methods). The two pulses are input into one erbium-doped fiber amplifier (EDFA) and amplified, then another two WDMs are used to control the pulses independently. This controlling includes time synchronization, polarization maintaining, filtering and intensity regulation. A free-space variable delay line, consisting of two lens couplers and their relative distance precisely adjusted in micrometers, is used to realize the time synchronization.

We use the resulting pair of precisely timed and linear polarized pulses to generate degenerate photon pairs with silicon photonics, which has been shown to be an efficient medium to generate various frequency non-degenerate two-photon and four-photon quantum states~\cite{Wang2018,Zhang2018,Silverstone2016,Sharping2006,Takesue2007,Takesue2008,Harada2008,Clemmen2009}. In the region labeled `generation' in Fig. 1(b), a Sagnac loop is shown, which is self-stabilized and a popular method for generating polarization entangled quantum states~\cite{Kim2006,Li2017,Fulconis2007,Takesue2008}. The interferometer includes one circulator, two half-wave plates (HWPs), two quarter-wave plates (QWPs), one polarization beam splitter (PBS) and the silicon waveguide. The silicon spiral nanophotonic waveguide placed inside the interferometer as the nonlinear material is 1~cm long and has a compact footprint ($\sim$170$\times$170 $\upmu$m$^2$). It has a propagation loss of 1\ dB and its cross-section is carefully designed and fabricated to be $\sim$450$\times$220 nm$^2$ to guarantee only fundamental transverse-modes supported. The waveguide also achieves near-zero dispersion in the wavelength band at approximately 1550\ nm~\cite{Zhang2018}, which greatly helps obtain a broadband and uniform SFWM gain spectrum and, thus, greatly increases the number of photon pairs generated. We use two identical grating couplers to couple in/out the pump and the generated photon pairs and they are precisely designed to support only TE mode coupling.

Before entering the Sagnac loop, the pump pulses are rotated to 45$^\circ$ linear polarization [i.e., 50\% horizontal (H) and 50\% vertical (V)] by adjusting the polarization controllers (PCs) carefully. In the Sagnac loop, these pump pulses are then split into the clockwise (H polarization) and counterclockwise directions (V polarization) by the PBS. The HWP and QWP in the Sagnac loop are carefully controlled so that the polarization states of the clockwise and counterclockwise pump pulses are both aligned to the horizontal axis (TE polarization) in the silicon waveguide. In this way, SFWM of $\rm{TE}_{p_1}+\rm{TE}_{p_2}\to\rm{TE}_s+\rm{TE}_i$ occurs with the same efficiency and time-correlated signal and idler photons are generated in each direction, where TE$_{p_1}$  and TE$_{p_2}$ represent the pump photon in different pump pulses and TE$_{s(i)}$ represents the signal (idler) photon. All photons are in TE polarization and no polarization rotation occurs in the light propagation along the silicon spiral waveguide. After being coupled out from the chip, the photon pairs in the clockwise and counterclockwise directions will be converted to the states of $\left|V_sV_i\right\rangle$ and $\left|H_sH_i\right\rangle$, respectively. The maximum polarization entangled two-photon state can be generated when the clockwise and anticlockwise SFWM in the nonlinear medium superpose together by the PBS and output from the Sagnac loop~\cite{Kim2006,Fulconis2007,Takesue2008}. This two-photon quantum state is expressed as
\begin{equation}\label{3}
  \left|\Phi\right\rangle=\frac{1}{\sqrt{2}}(\left|H_sH_i\right\rangle+\left|V_sV_i\right\rangle).
\end{equation}
In our work, the photon pairs have the same frequency, and the quantum state can be expressed in the number basis, that is, $\left|H_sH_i\right\rangle=\left| 20 \right\rangle_{HV}$ and $\left|V_sV_i\right\rangle=\left| 02 \right\rangle_{HV}$ and they are respectively the bi-photon states generated in the clockwise and counterclockwise directions, with $\left|nn'\right\rangle_{HV}$ denoting $n$ photons with polarization $H$ and $n'$ photons with polarization $V$ in the same spatial mode. In fact, when considering higher-order nonlinear terms, any $2n$ photon number entangled states can be generated in this configuration and expressed as (see Methods for more details)
\begin{equation}
\left|\Phi\right\rangle_{2n}=\frac{1}{n!}(\frac{1}{2}((a^\dagger_H)^2+(a^\dagger_V)^2))^n|00\rangle_{HV}.
\end{equation}

The generated states in the interferometer are extracted from the output port of the circulator, and then sent to an analysis stage, labeled as `characterization' in Fig. 1(b). A fiber filter with a bandwidth of $50$~GHz centered at $1550.11$~nm was used to remove the pump laser. Based on the energy conversion in SFWM, the filtering process prepares the desired degenerate photon pairs at the same time, which also have a central wavelength of $1550.11$~nm. The polarization and quantum correlation of the photon-pairs are kept unchanged after going through the fiber filter with the modulation by the PC. The three wave plates (two QWPs and one HWP) in the `characterization' region act as a full SU(2) gadget on orthogonally polarized modes~\cite{Campos1989}, which can be parameterized by two angles $\varphi$ and $\theta$ with
\begin{equation}\label{4}
  U(\varphi,\theta)=\left(\begin{matrix} \cos\theta & e^{i\varphi}\sin\theta \\ -e^{-i\varphi}\sin\theta & \cos\theta \end{matrix}\right),
\end{equation}
where the values of $\varphi$ and $\theta$ are determined by the orientation of the wave plates. The exact functional dependence of the angles is given in the Supplementary Section II. By combining this with a subsequent PBS and fiber beam splitters (FBSs), we can set the basis for tomography and interference measurements. Finally, the photons are detected using superconducting nanowire single-photon detectors (SCONTEL, dark count rate $100$~Hz, detector efficiency $85\%$ at C band). Note that the characterization configuration shown in Fig. 1(b) is only for the four-photon quantum state tomography measurement. For the two-photon quantum state, only one FBS is needed and we place two detectors after the FBS, as in the setup shown in ref.~\cite{Dieleman2017}. For the multi-photon interference measurement, sketches of the experimental setup are given in the Supplementary Fig. 4.

To verify the degenerate photon pair generation, we pump the silicon waveguide in a single direction and use one 50/50 FBS and two detectors to carry out coincidence measurements. By changing the time delay between the pump pulses, we observe the coincidence variation shown in Fig. 1(c). The dependence fits well with a Lorentzian function, which corresponds to the pump pulse pattern in our experiment. We then set the delay to $0$ and observe single-photon counts and two-photon coincidences as varying the pump power of one of the pulsed lasers. As shown in Supplementary Fig. 2, both single-photon counts and two-photon coincidences show linear profiles, which is significantly different from the single-pump case, where two-photon coincidences are expected to have a quadratic dependence with the pump power~\cite{Husko2013}. These results confirm the behavior of the dual-pump technique.

\subsection{B. Separated photon-pair generation}
\begin{figure}[t]
\centering
\includegraphics[width=7.0cm]{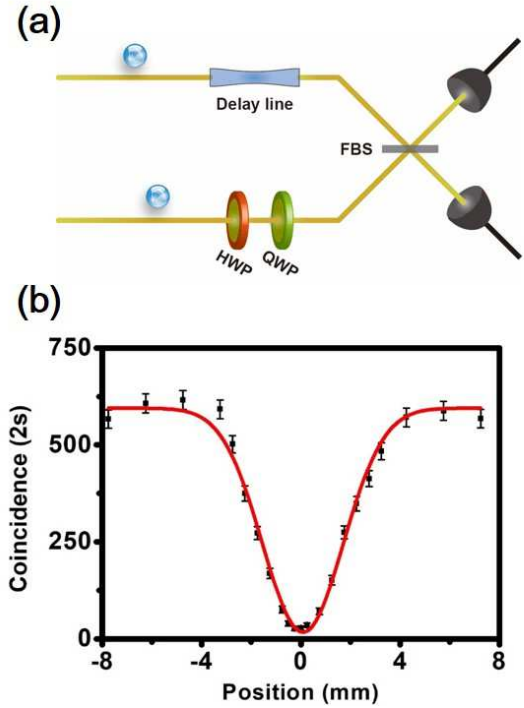}
\caption {\textbf{Hong-Ou-Mandel quantum interference.} \textbf{a}, A signal and idler photon are divided into different paths after the PBS in the characterization region of Fig. 1(b), one photon is delayed by a displacement and the other is polarization modulated so that the polarizations match. The two photons are then interfered on a fiber beam splitter (FBS). \textbf{b}, Two-photon coincidence counts as the relative time delay of the two photons is changed (no accidental coincidences subtracted). Error bars are $\pm\sqrt{N}$, based on the raw coincidences $N$.}
\end{figure}

The manipulated two-photon entangled state expressed in Eq. (1) can be used to prepare a separable two-photon state. By setting the angles to $\varphi=\pi/2$ and $\theta=\pi/4$, we can obtain the two-photon product state $\left|HV\right\rangle$. This generation process is the time-reversed version of Hong-Ou-Mandel (HOM) interference, as demonstrated in ref.~\cite{Silverstone2014,He2015,Chen2007}. With this state, we use the setup shown in Fig. 2(a) to perform a HOM type quantum interference measurement~\cite{Hong1987}. The signal and idler photons in different polarizations are split into different paths after the PBS in the characterization stage in Fig. 1(b). Then we send one photon to a tunable delay line and the other to a HWP and QWP. Thus, the optical path lengths and polarizations of the two photons can be matched precisely. The photon pairs are then interfered on a FBS and we record coincidences as we vary the delay line while keeping the polarization of the two photons completely consistent (Fig. 2(b)). The raw HOM interference visibility is measured as $97\pm2\%$ (net visibility $98\pm2\%$), which shows good indistinguishability between the signal and idler photons. Here, the visibility is defined as $V=(C_{\rm{max}}-C_{\rm{min}})/C_{\rm{max}}$, where $C_{\rm{max}}$ is the fitted maximum value and $C_{\rm{min}}$ is the fitted minimum value. For perfect quantum interference, $C_{\rm{min}}=0$ and $V=100\%$. The deviation of the visibility from unity may be due to the errors in the angles of the wave plates used in the state preparation.

\subsection{C. State characterization}
\begin{figure}[t]
\centering
\includegraphics[width=8cm]{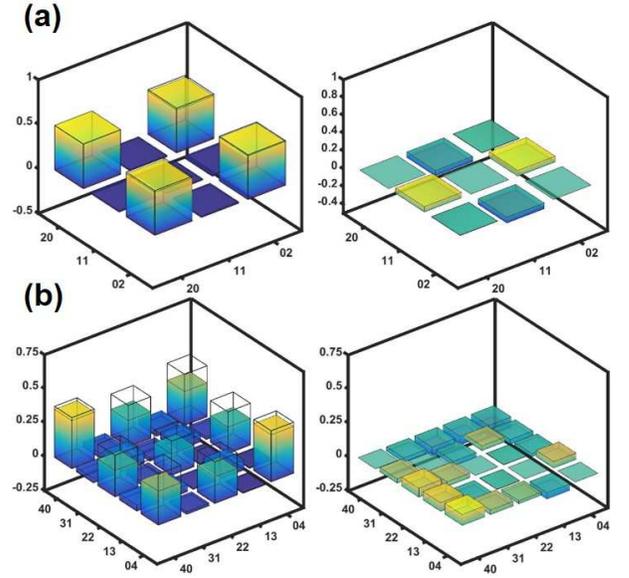}
\caption {\textbf{Quantum state tomography of frequency-degenerate entangled quantum states.} Experimental real (left column) and imaginary (right column) parts of the density matrices of \textbf{a}, two-photon maximally entangled state $\left|\Phi\right\rangle_2$, and \textbf{b},
four-photon polarization entangled state $\left|\Phi\right\rangle_4$. The states are reconstructed in the photon-number basis of two modes labelled as $\left|nn'\right\rangle$, where $n$ and $n'$ are the number of photons in $H$ and $V$ polarization, respectively. Solid frames represent the ideal cases.}
\end{figure}

\begin{figure}[t]
\centering
\includegraphics[width=8cm]{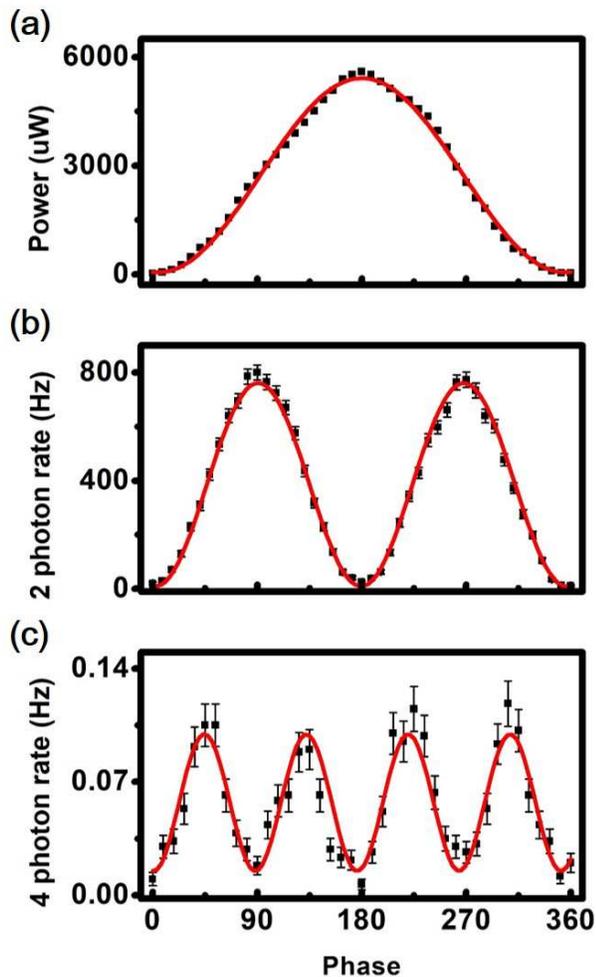}
\caption {\textbf{Interference measurement patterns for one, two and four photons.} \textbf{a}, Laser power as a function of the modulated phase angle with a CW laser input (equivalent to a one-photon pattern). \textbf{b}, Two-fold coincidence count rate pattern for the two-photon maximally entangled state $\left|\Phi\right\rangle_2$. \textbf{c}, Four-fold coincidence count rate pattern of one photon in $H$ polarization and three in $V$ polarization, resulting from the state $\left|\Phi\right\rangle_4$. Error bars are $\pm\sqrt{N}$, based on the raw coincidences $N$.}
\end{figure}

The previous section indirectly shows that we did produce two-photon entangled states and separated photon-pairs. In order to give a full characterization of the generated entangled quantum states, especially the four-photon entangled states, quantum state tomography is performed. The SU(2) gadget that the HWP and QWPs in the characterization stage shown in Fig.~1(b) represent, together with the PBS, was proposed in the theoretical work of ref.~\cite{Schilling2010} as a method to carry out state tomography on single-mode multi-photon polarization states, where one detector after the PBS is used to measure the $N$th-order intensity moment. In a real experiment, we can also use non-photon-number resolving detectors and beam splitters to achieve the same measurement, similar to the method demonstrated in characterizing multi-photon NOON states in recent experiments using free-space photons~\cite{Israel2012} and surface plasmons in waveguides~\cite{Dieleman2017,Chen2018}.

We first measure the two-photon maximally entangled state $\left|\Phi\right\rangle_2=\frac{1}{\sqrt{2}}(\left|20\right\rangle_{HV}+\left|02\right\rangle_{HV})$. The state tomography process needs data from 9 different wave plates settings, which are given in the Supplementary Table 1. We obtain the corresponding two-photon probability distribution of the state through coincidence measurements with one FBS and two detectors, with the dual-pump power of $P_1=P_2=80\ \upmu$W. We then use the maximum-likelihood-estimation method~\cite{James2001} to reconstruct the density matrix with this probability distribution, with the result displayed in Fig. 3(a). The fidelity between the measured density matrix $\rho_{\rm{exp}}$ and the ideal density matrix $\rho_{\rm{th}}$ is defined as $F={\rm Tr}(\rho_{\rm{exp}}\rho_{\rm{th}})$, where ${\rm Tr}$ is the trace. We obtain a raw fidelity of $0.98\pm0.06$, confirming that the generated quantum state is of high quality and very close to the ideal maximally entangled state. The errors are obtained from Monte Carlo calculations with the experimental data subject to Poisson statistics~\cite{James2001}. The deviation of the fidelity from unity may be due to the errors in the angles of the wave plates.

We then measure the four-photon polarization entangled state, obtained by pumping in both directions within the Sagnac loop at a higher power.
The four-photon entangled quantum state is given by
\begin{equation}\label{5}
\left|\Phi\right\rangle_4=\sqrt{\frac{3}{8}}(\left|40\right\rangle_{HV}+\left|04\right\rangle_{HV})+\frac{1}{2}\left|22\right\rangle_{HV}.
\end{equation}
Here the four-photon state we achieved has a same form as in ref.~\cite{Nagata2007}, although with a completely different approach. And in our approach, the integrated photonic source is based on a silicon waveguide placed in a Sagnac loop, which is more compact, more stable and uses less power.
The state tomography process for four-photon states needs data from 25 different wave plates settings, which are given in Supplementary Table 2. We obtain the probability distribution of the state through four-photon coincidence measurements with the experiment setup shown in the characterization region of Fig. 1(b). The density matrix reconstructed from the experimental data is displayed in Fig. 3(b), with the state having a raw fidelity of $0.72\pm0.07$. In this case, we set the dual-pump power at $P_1=P_2=400\ \upmu$W and the fiber filter bandwidth is set to $100$~GHz in order to increase the overall count rate. In addition, by pumping the silicon waveguide in a single direction, we can obtain multi-photon Fock states. We also performed the state tomography measurement for the four-photon Fock state and show the density matrix in the Supplementary Section III.

\subsection{D. Phase sensing}

The degenerate quantum states we have generated have broader application prospects than their non-degenerate counterparts. As an example, we briefly investigate the entangled states for use in super-resolution phase quantum metrology~\cite{Steuernagel2002,Sun2006,Nagata2007}. By setting the angle $\theta=\pi/4$ and varying $\varphi$ from $0$ to $2\pi$, we achieved two-fold and four-fold phase sensitivity, compared to the single-photon case. The results are shown in Fig. 4. We use a continuous wave (CW) laser set to diagonal polarization to provide a benchmark for the single-photon interference pattern (the single-photon case and classical CW case are the same), and the raw visibility is $98\pm2\%$. The raw visibilities of the two-photon and four-photon cases are $98\pm2\%$ and $74\pm5\%$, respectively. The integral time of each point in the four-photon interference measurements is $600$~s and the dual-pump power is $P_1=P_2=250\ \upmu$W. More details about the phase sensitivity measurement are provided in the Supplementary Section IV.

\section*{3. DISCUSSION}
Compared with the traditional free-space setup for the preparation of four-photon entangled states, as reported in ref.~\cite{Nagata2007}, where the state $\left|22\right\rangle_{HV}$ was used as an input and a complex interferometer was employed, our protocol is much more compact and simpler.  Only a fiber-based femtosecond laser is used and the pump power is just several hundreds of $\upmu$W, while the pump power for the free-space setups are usually several W.
Also, all the devices we used are communication compatible, and no extra spatial or temporal compensation processes are needed for multi-photon quantum states, which reduces the cost and technical difficulties in multi-photon quantum state generation.

Furthermore, the bulk optical elements used to manipulate polarization states in our experiment are fully correspondent with existing integrated optical elements used to manipulate path-encoded states~\cite{Bergamasco2017}, for example, the polarization states can be replaced by two waveguides associated with the $|1\rangle$ and $|0\rangle$ states and the PBS can be replaced by properly connecting the waveguides of the input ports to the waveguides of the output ports. Therefore, by replacing the Sagnac loop with a Mach-Zehnder interferometer, our frequency-degenerate multi-photon source could be further integrated on a single chip, which would greatly improve the source brightness and stability (see Supplementary Section V). The large-scale integration potential and stability of silicon photonics will also motivate the use of mulit-photon sources in practical applications.

The photon pairs we realized through time-reversed HOM interference could potentially constitute the foundation of many chip-scale degenerate quantum states, for example, Bell states and Greenberger-Horne-Zeilinger states. Therefore, our approach will be useful in on-chip operations for realizing quantum algorithms within integrated photonic circuitry. Very recently, frequency-degenerate photon pairs have also been realized directly in one silicon chip with a similar method, for studying boson sampling~\cite{Paesani2018}. This can be seen as another important application of frequency-degenerate multi-photon sources.

Note that the multi-photon entangled state demonstrated in our work is still not optimal, which is the part that should be improved first. Future work would start from the following aspects. The fiber filters we used have fixed bandwidths and center wavelengths, which are not perfectly matched between the pre-filters and post-filters, and using on-chip adjustable filters would be a better choice. Exploiting shorter pump lasers, which are more localized in energy, will also improve the quality of multi-photon states and fringe measurement. The four-photon rate should be further increased by improving the collection efficiency, e.g., the efficiency of the chip-fiber coupling and the post-filtering. For example, assuming that the collection efficiency in each channel is improved to 70\% (similar to that in free space~\cite{Wang2016}), the four-photon rate will increase to 150\ Hz, which will satisfy more practical applications. As a key component of the collection efficiency, the efficiency of the single grating coupler used in this work is about $-$5\ dB, which, fortunately, can be improved to, e.g., $-$0.9\ dB by introducing a special grating design~\cite{Marchetti2017}. This improvement makes a high collection efficiency truly achievable and, thus, offers an opportunity to achieve photonic sources with even more photons.

In summary, we have demonstrated an effective and convenient way to achieve a frequency-degenerate four-photon quantum state using silicon photonics. We demonstrated the possible use of the entangled states realized in our experiment for quantum metrology. Furthermore, this approach could be integrated on a single chip, where the stability and scalability of the integrated photonics setting makes it appealing for the preparation of more complex frequency-degenerate quantum sates. These states may then play an important role in the practical construction of quantum algorithms and related quantum information processes.

\section*{METHODS}

\noindent {\bf Pulse width estimation.} In our experiment, the pump pulse width is mainly determined by the fiber filters' bandwidth, which is 50\ GHz ($\sim$0.4\ nm). Because in the filtering process, the time-bandwidth product of the pulse should remain constant, the new pulse width can be calculated as $\tau_{new}=90\times10^{-3}\times80/0.4\ ps=18\ ps$. The Lorentzian function fitting of Fig. 1(c) shows a 3 dB bandwidth of 23 ps, a little higher than 18 ps. This is because the multiple fiber filters we used which have a small difference in bandwidth and central wavelength lead to the further reduction of pump pulse's bandwidth.

\noindent {\bf System efficiency.} We ascertained the efficiencies of all parts of our setup with laser light measurements. The silicon waveguide has a propagation loss of 1\ dB, and the grating coupler has a coupling loss of 5\ dB. The system for state manipulation and state measurement has a loss of 4.3\ dB. The post-filters have an inherent loss of 2\ dB and both detectors have an efficiency of 85\%\ ($-$0.7\ dB). The total loss is 12\ dB and needs to be reduced further for entangled state preparation with higher photon number.

\noindent {\bf Theory for photon pair generation.} The interaction Hamiltonian for the degenerate correlated photon pair generation by SFWM in each direction of the Sagnac loop is described as
\begin{equation}\label{6}
\widehat{H}_x=\frac{i\hbar\chi}{\sqrt{2}}((a^{\dagger}_x)^{2}-(a_x)^{2}),
\end{equation}
where $\hbar$ is the reduced Planck constant, $\chi$ is proportional to the third-order nonlinear susceptibility and the amplitude of the pump, and $a^{\dagger}_x\ (a_x)$ is the creation (annihilation) operator for correlated photon pairs emitted in the direction $x$. We assume that the pump light is strong and can be treated as a classical oscillator. The time evolution of the quantum state from the output port of the Sagnac loop when pumped in both directions is given by
\begin{equation}\label{7}
\left|\Psi\right\rangle=e^{-i(\widehat{H}/{\hbar})t}\left|00\right\rangle_{HV}.
\end{equation}
Here, we use $\widehat{H}$ to represent the total Hamiltonian in two directions defined as $\widehat{H}=\frac{1}{\sqrt{2}}(\widehat{H}_{H}+\widehat{H}_{V})$, and $\left|mn\right\rangle_{HV}$ represents the number states, where $m$ and $n$ are the numbers of photons emitted in the horizontal ($H$) and vertical ($V$) polarization, respectively. By using the disentangling theorem \cite{Walls2007,Harada2011}, the last equation can be re-expressed as follows,
\begin{equation}
\begin{aligned}
\left|\Psi\right\rangle&=\frac{1}{C^2}e^{{\Gamma}(\frac{1}{2}((a^{\dagger}_H)^{2})+(a^{\dagger}_V)^{2})}\left|00\right\rangle_{HV} \\
&=\frac{1}{C^2}(1+\widehat{H}'+\frac{1}{2!}\widehat{H}'^2+\frac{1}{3!}\widehat{H}'^3+\cdots)\left|00\right\rangle_{HV},
\end{aligned}
\label{8}
\end{equation}
where $\widehat{H}'=\Gamma(\frac{1}{2}((a^{\dagger}_H)^{2})+(a^{\dagger}_V)^{2}))$, $C=\mathrm{cosh}\chi t$ and $\Gamma=\mathrm{tanh}\chi t$. Therefore, any $2n$ photon number entangled state can be directly selected and expressed as
\begin{equation}
\left|\Phi\right\rangle_{2n}=\frac{1}{n!}(\frac{1}{2}((a^\dagger_H)^2+(a^\dagger_V)^2))^n|00\rangle_{HV}.
\end{equation}
For example, the two-photon entangled state
\begin{equation}\label{9}
\left|\Psi\right\rangle_2=\frac{1}{2}((a^{\dagger}_H)^{2}+(a^{\dagger}_V)^{2})\left|00\right\rangle_{HV}=\frac{1}{\sqrt{2}}(\left|20\right\rangle_{HV}+\left|02\right\rangle_{HV}),
\end{equation}
the four-photon entangled state
\begin{equation}\label{10}
\begin{aligned}
\left|\Psi\right\rangle_4&=\frac{1}{2!}(\frac{1}{2}((a^{\dagger}_H)^{2}+(a^{\dagger}_V)^{2}))^2\left|00\right\rangle_{HV} \\
&=\sqrt{\frac{3}{8}}(\left|40\right\rangle_{HV}+\left|04\right\rangle_{HV})+\frac{1}{2}\left|22\right\rangle_{HV},
\end{aligned}
\end{equation}
and the eight-photon entangled state
\begin{equation}\label{10}
\begin{aligned}
\left|\Psi\right\rangle_8&=\frac{1}{4!}(\frac{1}{2}((a^{\dagger}_H)^{2}+(a^{\dagger}_V)^{2}))^4\left|00\right\rangle_{HV} \\
&=\frac{\sqrt{70}}{16}(\left|80\right\rangle_{HV}+\left|08\right\rangle_{HV})+\\
&\frac{\sqrt{10}}{8}(\left|62\right\rangle_{HV}+\left|26\right\rangle_{HV})+\frac{3}{8}\left|44\right\rangle_{HV}.
\end{aligned}
\end{equation}

\noindent {\bf Source brightness estimation.} Here, we give a rough estimation of the source brightness based on the data in the fringe measurements. With the pump power $P_1=P_2=80\ \upmu$W, the maximum two-photon rate in the fringe measurement is around 800\ Hz, which corresponds to a generation rate about 201\ kHz, that is, 0.002 pair per pulse, when considering the total single photon loss 12\ dB. With the pump power $P_1=P_2=250\ \upmu$W, the photon pair generation rate is improved to $0.002\times250\times250/(80\times80)\approx0.02$ pair per pulse. Because the four-photon quantum state is generated by multiplexing two biphoton sources, its generation rate is squared compared to the biphoton state, that is, 0.0004 per pulse, 40\ kHz. At the same time, the four-photon event bears more postprocessing loss. For example, the single photon loss is 12\ dB, and the four-photon loss will be $12\times4=48$\ dB. Of course, the loss introduced by different experimental conditions in the four-photon fringe measurement should also be considered, such as the light splitting processing by the FBS, and measuring $\left|13\right\rangle_{HV}$ terms only, the four-photon count rate that can be detected will be around 0.06. This estimation value is very close to the experimental results in Fig. 4(c).

\noindent {\bf Optical apparatus setting.} In our experiment, the coupling angle was set at $10^\circ$ for fiber-chip coupling. The electrical signals from the detectors were collected and analyzed through a time-correlated single-photon counting (TCSPC) system, and the coincidence window was set as 0.8\ ns for two-photon state measurements. For four-photon state measurements, the electrical signals were analyzed by a UQD-Logic device and the coincidence window was set as 1\ ns.

\section*{DATA AVAILABILITY} Data are available from the authors upon reasonable request.

\section*{ACKNOWLEDGEMENTS} This work was supported by the National Key R \& D Program (No. 2016YFA0301700), the National Natural Science Foundation of China (NSFC) (Nos. 61590932, 11774333, 61725503, 61431166001), the Strategic Priority Research Program of the Chinese Academy of Sciences (No. XDB24030601), Anhui Initiative in Quantum Information Technologies (No. AHY130300), the Fundamental Research Funds for the Central Universities and the South African National Research Foundation SARChI programme. This work was partially carried out at the USTC Center for Micro and Nanoscale Research and Fabrication.

\section*{AUTHOR CONTRIBUTIONS}

All authors contributed extensively to the work presented in this paper. M.Z., Z.Y.Z. and D.X.D. provided the samples. L.T.F., Y.C. and X.F.R. performed the measurements, data analyses and discussions. M.L., H.L.R., G.P.G. and G.C.G. conducted theoretical analysis. L.T.F., M.T. and X.F.R. wrote the manuscript.

\section*{COMPETING INTERESTS}
The authors declare no competing interests.


\end{document}